\documentclass[twocolumn,showpacs,preprintnumbers,amsmath,amssymb]{revtex4}
\usepackage{bm}
\usepackage{amsmath}

\usepackage{graphicx}
\usepackage{dcolumn}

\begin{document}


\title{Observation of spontaneous path entanglement in the interference of quantum and classical light}
\title{Generation of multiphoton path entanglement by interference of quantum and classical light}
\title{Generation of scalable path entanglement by interference of quantum and classical light}
\title{Generation of multiphoton path entanglement for scalable phase super-sensitivity}

\title{Demonstration of a simple scheme for arbitrarily high photon-number path entanglement}
\title{Demonstration of high photon-number path entanglement in a scalable approach}
\title{Demonstration of multiphoton path entanglement by interference of quantum and classical light}
\title{Generating scalable number-path entanglement by interference of quantum and classical light}
\title{Scalable generation of  multiphoton path entanglement with high NOON-state fidelity}

\author{I. Afek}
\author{O. Ambar}%
\author{Y. Silberberg}%

\affiliation{%
Department of Physics of Complex Systems, Weizmann Institute of
Science, Rehovot 76100, Israel }
\date{\today}

\begin{abstract}
"Schr\textbf{$\ddot{\textbf{o}}$}dinger's cat" is a \textit{\textbf{gedankenexperiment}}
  intended to highlight conceptual difficulties in the interpretation of quantum mechanics \cite{SCHRODINGER35}. The generation of 'cat-like' states is central to numerous quantum information protocols. In particular, much attention has been drawn by states containing $\textbf{N}$ photons in a superposition of all being in one of two designated modes. These maximally path entangled states, known as 'NOON' states \cite{CONTEMPPHYS08DOWLING}, exhibit enhanced phase sensitivity and allow reaching the fundamental quantum limit of precision measurement\cite{PRA97OU}.  Such states could also be used for obtaining 'super-resolution' in quantum lithography \cite{PRL00}. Creation of NOON states in the lab,  has been limited to $\textbf{N=3}$ \cite{NAT04Steinberg,OE09}. Surpassing this has proven a formidable experimental challenge \cite{CONTEMPPHYS08DOWLING}. Here, we realize a scheme \cite{PRA07} for generation of high fidelity NOON states with arbitrarily large $\textbf{N}$.  We demonstrate the versatility of the scheme by measuring up to $\textbf{N=5}$ in a single setup, this is in contrast to previous experiments which were custom-designed for a specific $\textbf{N}$. The scheme is based on high order interference between 'quantum' down-converted light and 'classical' coherent light in the photon number basis. Our results verify the high degree of path entanglement which emerges naturally from the interference of two ubiquitous, easy to generate, states of light.
\end{abstract}

\maketitle

Entanglement is a distinctive feature of quantum mechanics which lies at the core of many new applications in the emerging science of quantum information \cite{COMPUTATIONBOOK}. These applications often require the use of large 'cat-like' states, composed of equal superpositions of  maximally different states. Realization of such  states could also allow experimental study of fundamental issues such as the effects of decoherence on many particle entanglement \cite{PRL08DAVIDOVICH,PRL04EIS}. Of particular interest are the maximally path entangled NOON states \cite{CONTEMPPHYS08DOWLING}, $|N::0\rangle_{a,b} \equiv \frac{1}{\sqrt{2}}  \left( |N,0\rangle_{a,b} +|0,N\rangle_{a,b}\right)$, which contain $N$  photons in an equal superposition of all being in one of two possible modes $a$ and $b$. The photons act as a collective entity acquiring phase at a rate $N$  times faster than a single photon with the same wavelength \cite{IntroQuantOPt}. Various schemes for NOON state generation have been suggested \cite{PRL09Kwiat,PRL07DOWLING,PRL07CABLE,PRA03WHITE,PRA02FIUR,PRA04HOF,CONTEMPPHYS08DOWLING}. However, the three photon record of leading experiments \cite{NAT04Steinberg,OE09} has proven difficult to surpass. We note that a 'NOON-like' four photon state has been generated, but only by using four rather than two spatial modes \cite{NAT04}. A number of experiments have used state projection to focus on the NOON component of various initial $N$  photon states \cite{PRA08OU,NAT04,PRL07,NJPH08}.

\begin{figure}
\includegraphics{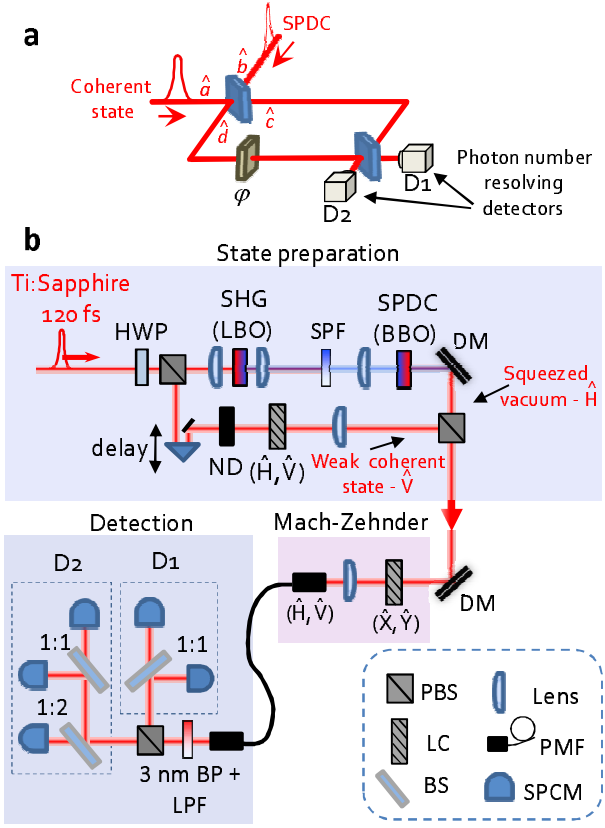}
\caption{\label{fig:setup} Experimental Setup for NOON state generation. \textbf{a}, Cartoon of the setup depicting a Mach-Zehnder (MZ) interferometer fed by a coherent state and SPDC. The NOON states occur in modes c and d after the first beamsplitter.  Measurement of multi-photon coincidences is performed using photon number resolving detectors. \textbf{b}, Detailed layout of the setup. A pulsed Ti:Sapphire oscillator with $120$ fs pulses @$80$ MHz is doubled using a $2.74$mm LBO crystal to obtain $404$nm  ultra-violet pulses with maximum power of $225$mW. These pulses then pump collinear degenerate type-I SPDC @$808$nm  using a $1.78$mm BBO crystal. The SPDC (H polarization) is coherently mixed with the coherent light (V polarization) using a polarizing beam-splitter (PBS). A thermally induced drift in the relative phase is corrected every few minutes using a LC (liquid crystal) phase retarder. The MZ  is polarization based in a collinear inherently phase stable design. The MZ phase is controlled using an additional LC phase retarder at $45$ degrees. The spatial and spectral modes are matched using a polarization maintaining fiber (PMF) and a $3$nm FWHM band-pass (BP) filter. Photon number resolving detection is performed using an array of single photon counting modules (SPCM, Perkin Elmer). Additional components: long-pass filter (LPF), short-pass filter (SPF) and dichroic mirror (DM).}
\end{figure}

In this letter we experimentally realize a scheme introduced by Hofmann and Ono \cite{PRA07}, which yields high fidelity NOON states for arbitrary N.  The approach is experimentally appealing due to its inherent simplicity relying on a fundamental unmodified multiphoton interference effect. Consider a $50/50$ beam-splitter fed by a coherent state, $|\alpha\rangle_{a}$, in one input port and spontaneous parametric down-conversion (SPDC), $|\xi\rangle_{b}$, in the other (see Fig. \ref{fig:setup}a). The input states are defined in the conventional way \cite{IntroQuantOPt}
This property leads to enhanced phase sensitivity which can be used for reaching the fundamental Heisenberg sensitivity limit \cite{CONTEMPPHYS08DOWLING} and sub-wavelength lithography \cite{PRL00}. Scaling up the size of experimental NOON states is therefore of practical as well as fundamental interest.

Various ingenious schemes for NOON state generation have been suggested \cite{PRL09Kwiat,PRL07DOWLING,PRA03WHITE,PRA02FIUR,PRA04HOF}. These schemes, however are typically quite elaborate making it difficult to overcome the three photon record of leading experiments \cite{NAT04Steinberg,OE09}. We note that a 'NOON-like' four photon state has been generated, but only by using four rather than two spatial modes \cite{NAT04}. A number of experiments have used state projection to focus on the NOON component of various initial $N$  photon states \cite{PRA08OU,NAT04,PRL07,NJPH08}.

\begin{figure}
\includegraphics{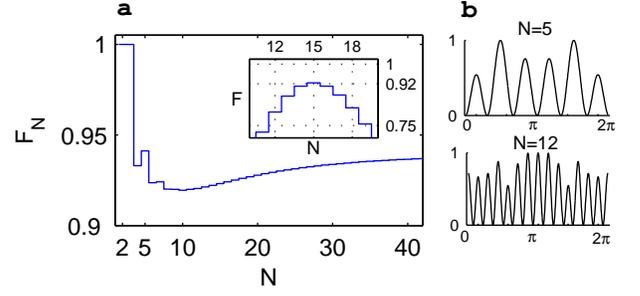}
\caption{\label{fig:fidelity} The theoretical fidelity of the generated NOON states \textbf{a}, Fidelity, $F_N$ vs. $N$ in an ideal setup. The pair amplitude ratio $\gamma$, which maximizes the NOON state overlap was chosen separately for each $N$. The fidelity approaches $0.943$ asymptotically for large $N$. The inset shows the fidelity for nearby $N$ when $\gamma$ is optimized for $N=15$. In this case $F_N>0.75$ for $N=12$ to $19$, simultaneously.  \textbf{b}, Simulated $N$-fold coincidences as a function of Mach-Zehnder phase for $N=5,12$ demonstrating N-fold super-resolution. }
\end{figure}

In this letter we experimentally realize a scheme introduced by Hofmann and Ono \cite{PRA07}, which yields high fidelity NOON states for arbitrary N.  The approach is experimentally appealing due to its inherent simplicity relying on a fundamental unmodified multiphoton interference effect. Consider a $50/50$ beam-splitter fed by a coherent state, $|\alpha\rangle_{a}$, in one input port and spontaneous parametric down-conversion (SPDC), $|\xi\rangle_{b}$, in the other (see Fig. \ref{fig:setup}a). The input states are defined in the conventional way \cite{IntroQuantOPt}
\begin{eqnarray}
| \alpha \rangle &=& \sum_{n=0}^{\infty} e^{-\frac{1}{2}|\alpha|^2} \frac{\alpha^n}{\sqrt{n!}}|n\rangle, \quad \alpha = |\alpha|e^{\imath \phi_{cs}}
\notag\\
\ | \xi \rangle &=& \frac{1}{\sqrt{\cosh r }}\sum_{m=0}^{\infty}(-1)^m \frac{\sqrt{(2m)!}}{2^m m!} \notag \\
&{}& \quad \quad \quad \quad \quad \quad \quad \quad  \times(\tanh r)^m |2m\rangle.\label{eq:spdc_def}
\end{eqnarray}
Where the phase of $|\xi\rangle$ has been set arbitrarily to zero leaving the relative phase of the two inputs to be determined by $\phi_{cs}$.
We denote the pair amplitude ratio of the coherent state and SPDC inputs by $\displaystyle \gamma = |\alpha|^2 / r$.
The output state, $|\psi_{out}\rangle_{c,d}$, is highly path entangled. A general $N$ photon two-mode state can be written as $\sum_{k=0}^{N}u_k|k\rangle_c|N-k\rangle_d$. The creation of an ideal NOON state would require elimination of all the non-NOON components i.e. $u_1,\ldots ,u_{N-1}=0$. Remarkably the present scheme does this almost perfectly using the naturally emerging multiphoton interference. The fidelity of the output state's normalized $N$  photon component with a NOON state is  $F_N\equiv| \langle N::0 | \psi_{out}^{N} \rangle|^2$. It can be shown that by choosing  $\phi_{cs}=\pi$ and optimizing $\gamma$ for each N, one can achieve $F_N > 0.92$ for arbitrary N, see  Fig. \ref{fig:fidelity}a. The theoretical overlaps for the states generated in this work are  $F_N=1,1,0.933,0.941$ for $N=2,3,4,5$ respectively. We note that the four photon value is much higher than the theoretical fidelity of $0.75$ obtainable using down-conversion only \cite{SCI07,NJPH08}.

\begin{figure*}[ht]
\includegraphics{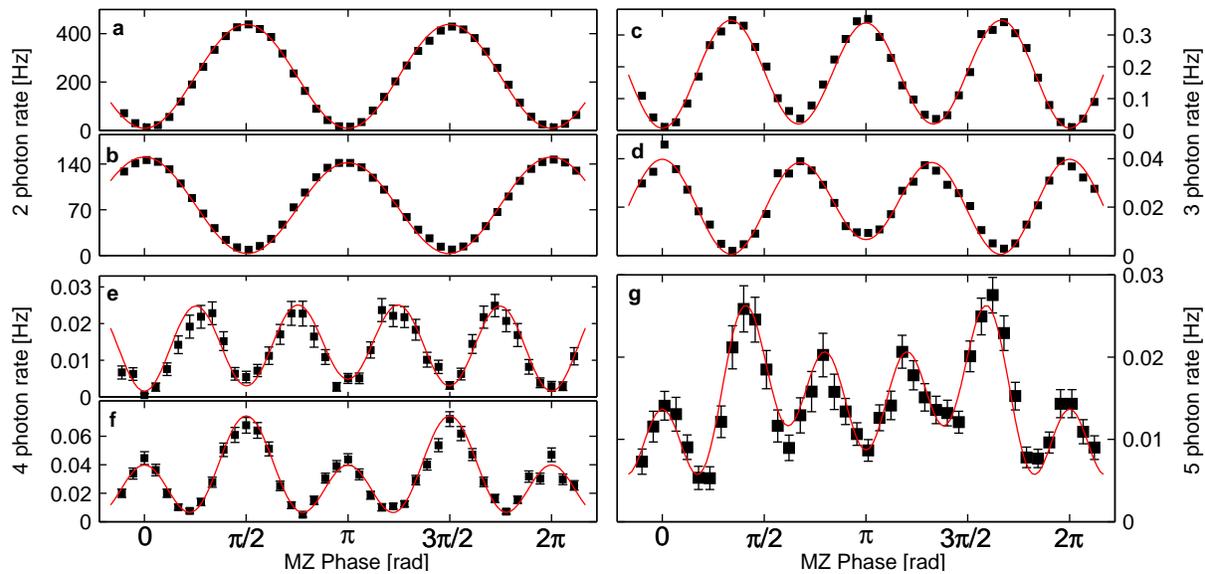}
\caption{\label{fig:coincidence} Experimental results. Coincidence measurements demonstrating $N$-fold super-resolution for $N=2,3,4$ and $5$ with no background substraction. Error bars indicate $\pm \sigma$ statistical uncertainty. The number of 'clicks' in detector $D_{1,(2)}$ is denoted $N_{1,(2)}$  (see Fig. \ref{fig:setup}). Two photon rate with \textbf{a}, $N_{1},N_{2} = 1,1$ and \textbf{b}, $N_{1},N_{2} = 0,2$. Three photon rate with \textbf{c}, $N_{1},N_{2} = 2,1$  and \textbf{d}, $N_{1},N_{2} = 0,3$. Four photon rate with  \textbf{e}, $N_{1},N_{2} = 3,1$  and \textbf{f}, $N_{1},N_{2} = 2,2$. Five photon rate with \textbf{g}, $N_{1},N_{2} = 2,3$. Solid lines are obtained from an analytical model. For each $N$ the pair amplitude ratio $\gamma$, was chosen separately to maximize the NOON state fidelity. The optimal values (obtained analytically) are $\gamma_{2}=\gamma_{3}=1$, $\gamma_{4}=\sqrt{3}$ and $\gamma_{5}=9/\left( \sqrt{10}+1 \right) \approx 2.16$. Visibility of the sinusoidal patterns with $N$ oscillations is determined by a weighted least squares $\sin,\cos$ decomposition restricted to frequencies of $0, 1,\ldots, N$. The values for plots \textbf{a} \-- \textbf{g} are $V=(95 \pm 0.0)\%$, $(88 \pm 0.03)\%$, $(86\pm0.6)\%$, $(80 \pm 1.9)\%$, $(74\pm3)\%$, $(73\pm2.4)\%$ and $(42\pm 2)\%$ respectively.}
\end{figure*} 

 The experiment (see Fig. \ref{fig:setup} for details) involves the generation of SPDC and coherent light with the same spatial and spectral modes.
 The beams are prepared in perpendicular linear polarizations ($H,V$) and spatially overlapped using a polarizing beam-splitter (PBS). A Mach-Zehnder is then implemented in an inherently phase stable, collinear design so that the NOON  state mode subscripts $c,d$ may now be replaced by $H,V$.
 After application of the Mach-Zehnder (MZ) phase shift $\varphi$, the state $|N0\rangle_{H,V} + |0N\rangle_{H,V}$ evolves to \cite{CONTEMPPHYS08DOWLING} $|N0\rangle_{H,V} + e^{\imath N\varphi} |0N\rangle_{H,V} $. This gives rise to interference oscillations $N$ times faster than those of a single photon with the same frequency.

 The experimental results are shown in Fig. \ref{fig:coincidence}. Two to five photon coincidence rates were measured as a function of the MZ phase and are shown with no background substraction. We denote the number of photon 'clicks' in $D_1(D_2)$ by $N_1(N_2)$. The $N_1,N_2$ coincidence rate is expected to demonstrate a de-Broglie wavelength \cite{IntroQuantOPt} of $\lambda/N$, where  $N=N_1+N_2$. The red curves are produced by an analytical model of the experiment, accounting for the overall transmission and the positive operator-value measures (POVM) of the detectors. The two and three photon results were taken simultaneously with $\gamma_2=\gamma_3=1$ where the subscript of $\gamma$ denotes the value of $N$ for which it was optimized. For this measurement we used a mere $2$mW of ultraviolet power to pump the SPDC.  The results in Figs. \ref{fig:coincidence}c,\ref{fig:coincidence}d constitute the highest visibility measurement of a three photon NOON state to date \cite{NAT04Steinberg,OE09} ($86\pm0.6\%$ for $2,1$ and $80\pm1.9\%$ for $3,0$). The four photon measurements were taken with $\gamma_4 = \sqrt{3}$ and using $25$mW of ultraviolet pump. They exhibit similar visibilities for both the $2,2$ $(74\pm3)\%$ and the $3,1$ $(73\pm2.4)\%$ possibilities which fits well with our theoretical predictions. Finally, the $N=5$ result is, to the best of our knowledge, the first realization of a five photon NOON state ($V=42\pm2\%$ for $3,2$). This measurement was taken using $215$mW of ultraviolet pump and setting $\gamma_5=9/(\sqrt{10}+1)\sim2.16$ implying that  $(\gamma_{5}^2\sim)4.7$ times more photon pairs originate from the coherent state than from the SPDC. The above visibilities are incompatible with any known classical model.

The realized scheme has several unique properties \cite{PRA07}. Most importantly, it works naturally for arbitrary $N$. This requires no alterations to the setup except for use of detectors which can resolve $N$  photon events and setting $\gamma$ appropriately. This is in contrast to previous experiments which were customized to a specific value of $N$ \cite{SCI07,NAT04,NAT04Steinberg,OE09}. In addition, most of the photons in this scheme originate from the coherent (classical) light source which is practically unlimited in intensity. This eliminates the need for bright SPDC sources, providing a significant experimental simplification. In fact, it can be shown that  $\gamma_N \sim N/2$ for large $N$ where $\gamma_N$ is the optimal pair amplitude ratio for a given $N$. Thus, the coherent state's two photon probability is approximately $N^2/4$ times higher than that of the  SPDC i.e. the higher the value of $N$ the larger the ratio of classical to quantum resources. Finally, the scheme involves no state projection or post selection, implying that all the $N$-photon events contribute to the measurable $N$-photon interferences.

It is interesting to note that the $N$-fold coincidence plots of Fig. \ref{fig:fidelity}b exhibit $N$ zero points as expected by perfect maximally path entangled states albeit with somewhat modulated peak heights. In fact, it has been shown \cite{PRA07} that due to the high fidelity we can expect a phase sensitivity that is only slightly lower than the Heisenberg limit for a given $N$ photon component. Thus, the small deviation from an ideal NOON state is not a major limiting factor and is more than compensated for by the resulting intrinsic $100\%$ efficiency. Furthermore, while the state at hand is optimized for a given $N=N_0$, there is a range of $N$ values surrounding $N_0$ which also have considerable NOON fidelity (see inset of Fig. \ref{fig:fidelity}a). Thus, the generated state is actually a superposition of NOON states with various photon numbers, each of which contributes to the enhanced phase sensitivity. As a result, the current method has the potential for beating the standard quantum limit both for specific $N$ and while accounting for the complete photon number distribution \cite{PRL08,NJPH08}.

The visibility of the experimental plots in Fig. \ref{fig:coincidence} is determined by the overall setup transmission which we denote $\eta$. For $\eta<1$, higher order events contribute a background to the $N$-fold coincidence rate. In the current setup we estimate $\eta\sim0.12$ based on the SPDC coincidence to singles ratio (see supplementary material for details). We note that since measuring larger states with this scheme does not require a brighter SPDC source, the only limiting factor in this experiment is the overall transmission, $\eta$. Simulations show that assuming an overall transmission of $\eta=0.5$,  de-Broglie wavelengths of up to $\lambda/9$ are readily observable even with the current, relatively modest, SPDC flux (see supplementary material). In this respect, our experiment highlights the need for high purity SPDC sources which can be spectrally mode matched to a coherent state. Improved transmission can be obtained by way of SPDC generation in separable spectral modes \cite{PRL08Lundeen,NJPHYS08WALMSLEY,PRL07Torres} allowing removal of the $3$nm band-pass filter. In addition,  improved single mode coupling of the photon pairs and use of high efficiency photon number resolving detectors \cite{CP07,OE08} are instrumental.

%


It is interesting, that a setup very similar to the one in Fig. \ref{fig:setup}a, yet with much stronger light fields, is commonly used for obtaining quantum noise reduction using homodyne detection \cite{IntroQuantOPt}. Homodyne detection is a highly developed technique, based on the measurement of the continuous variable field quadratures. Our experiment demonstrates that extending the concept of squeezed vacuum and coherent light interference into the weak local oscillator regime is extremely fruitful. The states emerging from this interference, for a given N, are almost perfect NOON states. This implies a fundamental connection between quantum noise reduction (with continuous variables) and creation of NOON states (in photon-number resolving experiments). We note that two photon interference between coherent light and down-conversion has been previously observed \cite{PRA06PIT,PRL01OU}.

In conclusion, we have implemented a scheme for generation of path entangled states with high NOON state overlap for arbitrary $N$. The simplicity of the approach allows probing maximal path entanglement at previously unaccessible photon numbers. States of up to five photons have been obtained in a single setup. The scheme is conceptually simple and generalizes naturally to any $N$, requiring no post-selection or state projection. We expect this experiment will pave the way for observation of even larger states, thus helping expand the frontiers of experimental quantum metrology.
\begin{acknowledgments}
We thank Barak Dayan for stimulating discussions. Itai Afek gratefully acknowledges the support of the Ilan Ramon Fellowship.
Financial support of this research by the German Israeli
foundation is gratefully acknowledged.
Correspondence and requests for materials
should be addressed to Afek. I ~(email: itai.afek@weizmann.ac.il).
\end{acknowledgments}


\end{document}